\documentclass[fleqn,10pt]{wlscirep}

\title{Evidence of $s$-wave superconductivity in the noncentrosymmetric La$_7$Ir$_3$}

\author[1]{B. Li}
\author[2,3]{C. Q. Xu}
\author[2]{W. Zhou}
\author[4]{W. H. Jiao}
\author[5,6]{R. Sankar}
\author[2]{F. M. Zhang}
\author[2]{H. H. Hou}
\author[2]{X. F. Jiang}
\author[2]{B. Qian}
\author[7]{B. Chen}
\author[8]{A. F. Bangura}
\author[2,3,*]{Xiaofeng Xu}
\affil[1]{College of Science, Nanjing University of Posts and Telecommunications, Nanjing 210023, China}
\affil[2]{Advanced Functional Materials Lab and Department of Physics, Changshu Institute of Technology, Changshu 215500, China}
\affil[3]{Department of Physics, Hangzhou Normal University, Hangzhou 310036, China}
\affil[4]{Department of Physics, Zhejiang University of Science and Technology, Hangzhou 310023, China}
\affil[5]{Institute of Physics, Academia Sinica, Nankang, Taipei R.O.C. Taiwan 11529}
\affil[6]{Center for Condensed Matter Sciences, National Taiwan University, Taipei 10617, Taiwan}
\affil[7]{Department of Physics, University of Shanghai for Science $\&$ Tehcnology , Shanghai, China}
\affil[8]{Max-Planck-Institut f\"{u}r Festk\"{o}rperforschung, Heisenbergstr. 1, D-70569 Stuttgart, Germany}
\affil[*]{xiaofeng.xu@cslg.edu.cn}

\date{\today}

\begin{abstract}
Superconductivity in noncentrosymmetric compounds has attracted sustained interest in the last decades. Here we present a detailed study on the transport, thermodynamic properties and the band structure of the noncentrosymmetric superconductor La$_7$Ir$_3$ ($T_c$ $\sim$2.3 K) that was recently proposed to break the time-reversal symmetry. It is found that La$_7$Ir$_3$ displays a moderately large electronic heat capacity (Sommerfeld coefficient $\gamma_n$ $\sim$ 53.1 mJ/mol $\text{K}^2$) and a significantly enhanced Kadowaki-Woods ratio (KWR $\sim$ 32 $\mu\Omega$ cm mol$^2$ K$^2$ J$^{-2}$) that is greater than the typical value ($\sim$ 10 $\mu\Omega$ cm mol$^2$ K$^2$ J$^{-2}$) for strongly correlated electron systems. The upper critical field $H_{c2}$ was seen to be nicely described by the single-band Werthamer-Helfand-Hohenberg model down to very low temperatures. The hydrostatic pressure effects on the superconductivity were also investigated. The heat capacity below $T_c$ reveals a dominant $s$-wave gap with the magnitude close to the BCS value. The first-principles calculations yield the electron-phonon coupling constant $\lambda$ = 0.81 and the logarithmically averaged frequency $\omega_{ln}$ = 78.5 K, resulting in a theoretical $T_c$ = 2.5 K, close to the experimental value. Our calculations suggest that the enhanced electronic heat capacity is more likely due to electron-phonon coupling, rather than the electron-electron correlation effects. Collectively, these results place severe constraints on any theory of exotic superconductivity in this system.
\end{abstract}

\begin{document}                              
\flushbottom
\maketitle
\thispagestyle{empty}


\section*{Introduction}
The discovery of superconductivity in the noncentrosymmetric compounds, i.e., crystals possessing no inversion center, has generated immense interest in condensed matter physics\cite{Yuan17}. The absence of inversion symmetry and the associated antisymmetric spin-orbit coupling (ASOC) allow the admixture of spin-singlet and spin-triplet components in the order parameter which is otherwise highly unlikely in centrosymmetric materials due to parity conservation. As a result, noncentrosymmetric superconductors (NCSs) often exhibit peculiar superconducting properties, such as the violation of the Pauli paramagnetic limit and the presence of nontrivial line or point nodes in the order parameter.

The material realization of the NCS was reported in the heavy-fermion compound CePt$_3$Si~\cite{Bauer2004,Bauer2005,Bauer2007} in which superconductivity coexists with antiferromagnetic order. Many other notable NCSs including Ca(Ir,Pt)Si$_3$,~\cite{Eguchi2011} La(Rh,Pt,Pd,Ir)Si$_3$,~\cite{Anand2011,Smidman2014,Anand2014} Li$_2$(Pd,Pt)$_3$B,~\cite{Togano2004,Yuan2006,LiPdPtB} LaNiC$_2$,~\cite{Pecharsky1998,Hillier2009,Bonalde2011}, Re$_6$Zr \cite{Matthias,Singh2014}, PbTaSe$_2$ \cite{Cava14,Xu_PbTaSe} etc., were also discovered. Among these NCSs, those with broken time reversal symmetry (TRS) were extremely rare and as such have attracted broad attention from both theorists and experimentalists. The broken TRS implies the spontaneous magnetization either due to nonunitary triplet pairing or from chiral singlet states such as $d$+$id$.

Recent muon spin relaxation ($\mu$SR) measurements on the polycrystalline sample of a noncentrosymmetric superconductor La$_7$Ir$_3$ ($T_c$ $\sim$2.3 K) reveal spontaneous static or quasistatic magnetic fields, suggesting the breaking of TRS in its superconducting state \cite{Baker2015,Biswas}. This implies that La$_7$Ir$_3$ may be a candidate for unconventional superconductor as suggested. Further transverse $\mu$SR measurements revealed an isotropic $s$-wave gap with enhanced electron-phonon coupling. However, other properties of this putative TRS-breaking state have never been reported thus far. In this context, it is necessary to study the transport and thermodynamic properties of this intriguing state and search for the evidence of TRS breaking state. For example, the triplet pairing with broken TRS often leads to an upper critical field far above the Pauli paramagnetic limit and has nodal or anisotropic gap functions.

In this study, we explored the noncentrosymmetric La$_7$Ir$_3$ superconductor via ultra-low temperature transport and thermodynamic measurements, complementary with the first principles calculations. Remarkably, the upper critical field in this system is lower than the Pauli limit and can overall be described by the one-band Werthamer-Helfand-Hohenberg (WHH) model. A moderately large electronic specific heat $\gamma_n$ $\sim$ 53.1 mJ/mol $\text{K}^2$ was seen in the normal state and quasiparticle excitations in the superconducting state can be well characterized by a single $s$-wave gap with the magnitude of $2\Delta_g/k_B T_c = 3.56$, very close to the BCS value of 3.5. Besides, the Hall effect, high pressure effect and the band structure were also investigated, which consistently support a phonon-mediated BCS pairing scenario for La$_7$Ir$_3$.

\begin{figure}
\begin{center}
\includegraphics[width=8.2cm]{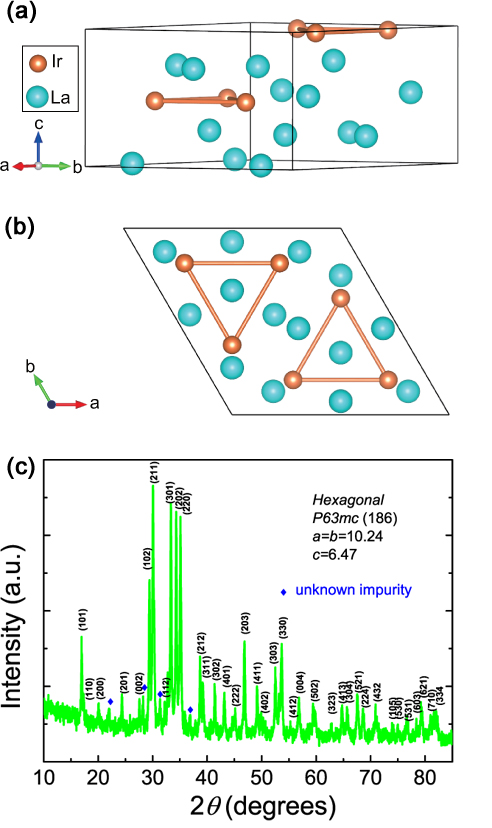}
\caption{\label{XRD} (a) Crystal structure of La$_{7}$Ir$_{3}$. Optimized Wyckoff position of Ir is 6$c$(0.81, 0.19, 0.47). Optimized Wyckoff positions of three non-equivalent La atoms are 2$b$(1/3, 2/3, 0.76), 6$c$(0.46, 0.54, 0.22) and 6$c$(0.87, 0.13, 0.02), respectively. (b) The structure projected on the $ab$-plane. (c) Powder x-ray diffraction (XRD) pattern of La$_7$Ir$_3$.}
\end{center}
\end{figure}

\section*{Results and Discussion}

\begin{figure}
\begin{center}
\includegraphics[width=8.0cm]{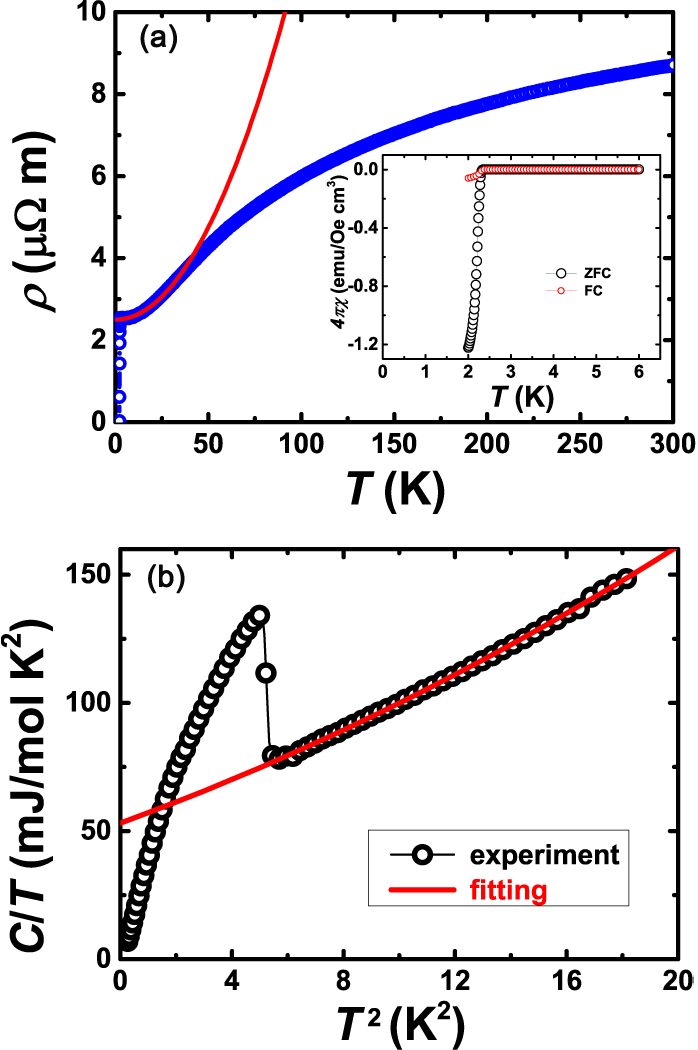}
\caption{\label{RTCT} (a) Temperature dependence of resistivity $\rho$ of La$_7$Ir$_3$. The red solid line is a fit based on the Fermi-liquid description, i.e. $\rho(T) = \rho_0 + AT^2$. The fitting obtained residual resistivity $\rho_0$ is 2.5 $\mu\Omega$ m, and transport coefficient $A$ is $9\times10^{-2}$ $\mu\Omega$ cm $\text{K}^{-2}$. Inset shows the temperature dependence of magnetization for both zero-field cooling (ZFC) and field cooling (FC) processes. (b) Temperature dependence of specific heat plotted as $C/T$ versus $T^2$. The red solid line represents the fit to the normal state $C$ based on the equation $C/T = \gamma_n + \beta_n T^2 + \alpha_n T^4$. The obtained parameters are $\gamma_n = 53.1$ mJ/mol $\text{K}^2$, $\beta_n = 3.96$ mJ/mol $\text{K}^4$ and $\alpha_n = 0.072$ mJ/mol $\text{K}^6$, respectively.}
\end{center}
\end{figure}

\begin{table}
\begin{center}
\caption{\label{Coefficient} Transport and thermodynamic parameters for some well-known compounds \cite{Analytis,Mun,Bauer2004}.}
\begin{tabular}{ccccc}
Comp. & $A$ & $\gamma_n$ & KWR \\
\ & $\mu\Omega$ cm $\text{K}^{-2}$ & mJ/mol $\text{K}^2$ & $\mu\Omega$ cm mol$^2$K$^2$J$^{-2}$\\
\hline
CePt$_3$Si  & 2.35 & 390 & 15.5\\
BaFe$_2$As$_{2-2x}$P$_{2x}$  & 0.009 & 7 & 7 - 15\\
PtSn$_4$  & 2$\times 10^{-4}$ & 4 & 12.5\\
La$_7$Ir$_3$ & 0.09 & 53.3 & 32\\
\end{tabular}
\end{center}
\end{table}

\begin{figure}
\begin{center}
\includegraphics[width=8.0cm]{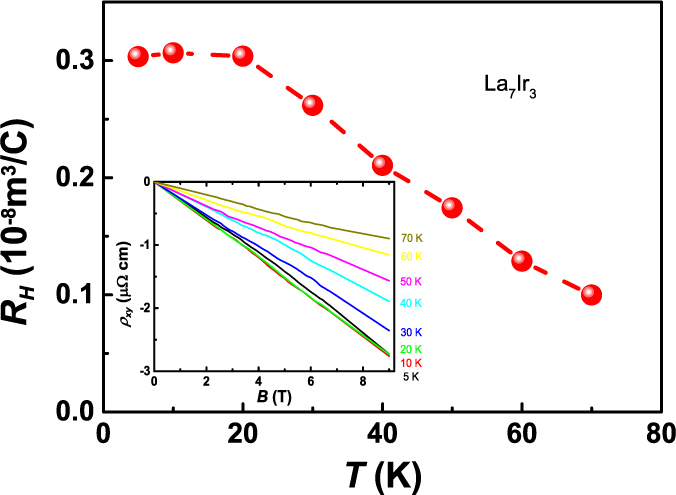}
\caption{\label{hall} Temperature dependence of Hall coefficient $R_H$ of La$_7$Ir$_3$. Inset: Magnetic field dependence of Hall resistivity $\rho_{xy}$ at different temperatures.}
\end{center}
\end{figure}

\begin{figure}
\begin{center}
\includegraphics[width=8.0cm]{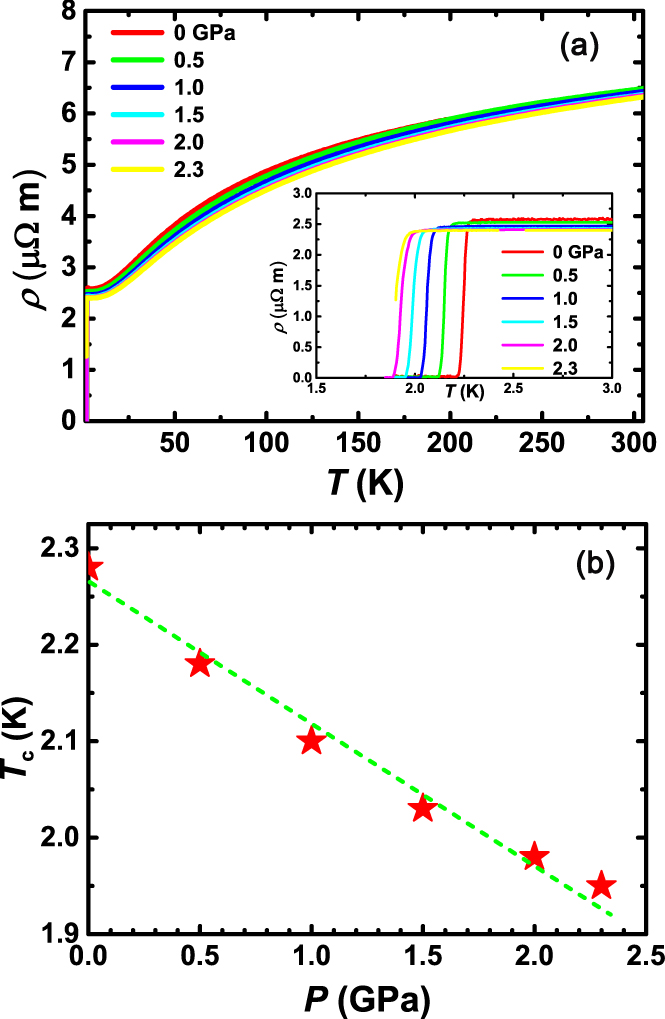}
\caption{\label{HP} (a) Temperature dependence of resistivity at different pressures. Inset is an enlarged view of the resistivity at low temperatures. (b) Pressure dependence of the superconducting transition temperature $T_c$. The green dashed straight line is a guide to the eye.}
\end{center}
\end{figure}

\begin{figure*}
\begin{center}
\includegraphics[width=13cm]{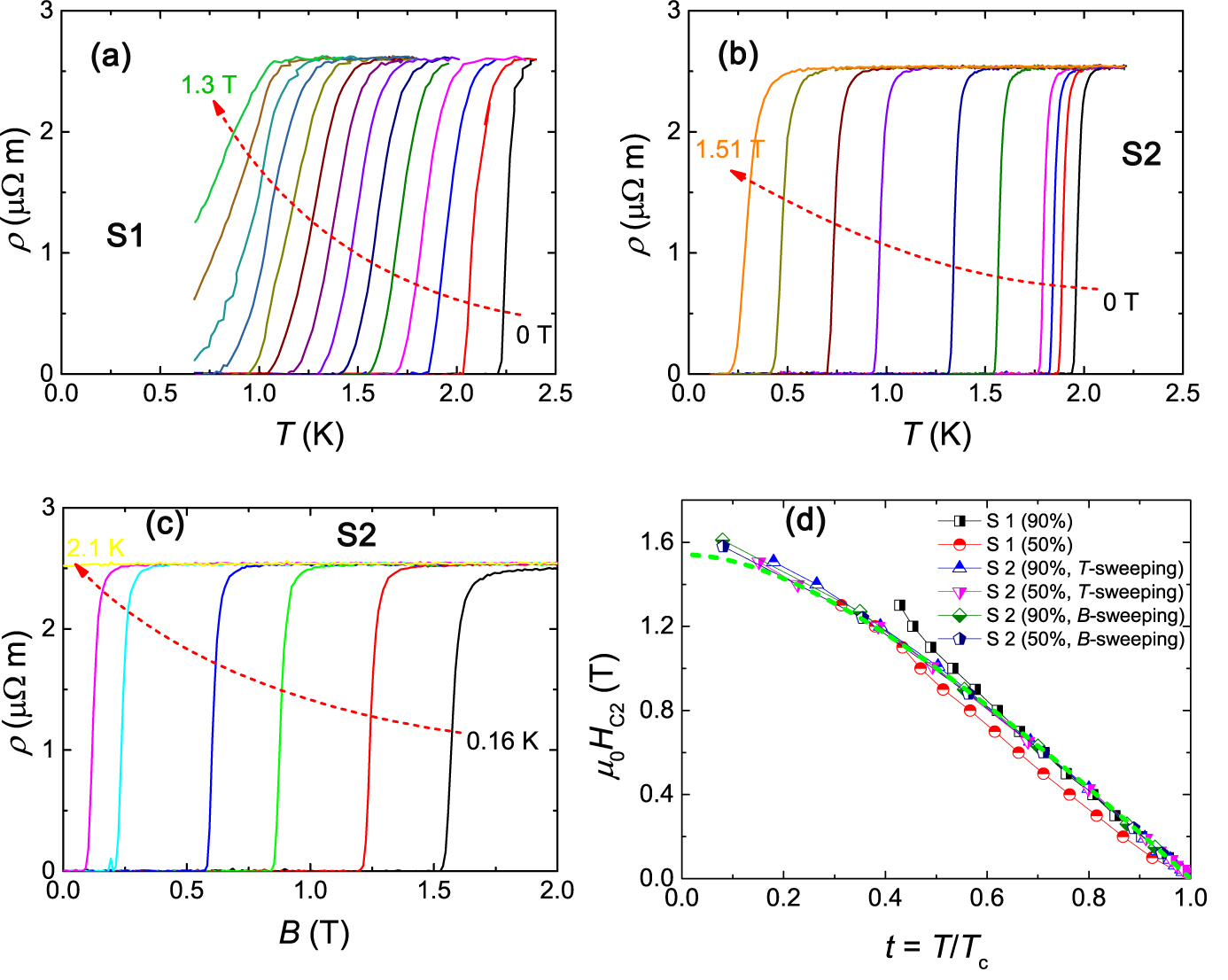}
\caption{\label{Hc2} (a) Temperature dependence of resistivity measured using a $^3$He system at different magnetic fields. (b) and (c) show the resistive superconducting transitions measured in a dilution refrigerator with the fixed-field temperature sweeps and the fixed-temperature field sweeps, respectively. (d) Normalized temperature ($t = T/T_c$) dependence of upper critical fields ($\mu_0H_{c2}$). 90\% and 50\% mean the criteria used to obtain the $\mu_0H_{c2}$ values, defining the data point of 90\% and 50\% of the normal state resistivity $\rho_n$ just above $T_c$. The green dashed line represents the WHH fitting without considering the spin-paramagnetic effect and the spin-orbit interaction ( ($\alpha = 0$, $\lambda = 0$).}
\end{center}
\end{figure*}

The schematic view of the crystal structure of La$_{7}$Ir$_{3}$ is shown in Fig. \ref{XRD}. La$_{7}$Ir$_{3}$ crystallizes in a hexagonal Th$_{7}$Fe$_{3}$ structure with the space group $P6_{3}mc$ (No.186). Its structure consists of alternate stacking of iridium monolayer and lanthanum bilayer along the $c$-axis. We optimized the coordinates of atoms with the experimental lattice parameters $a$=$b$=10.2376$\AA$, $c$=6.4692$\AA$ \cite{Baker2015}. Fig. \ref{RTCT} (a) shows the temperature dependent resistivity of La$_7$Ir$_3$. A sharp superconducting transition is observed around 2.3 K, which is consistent with the diamagnetization measurement shown in the inset of Fig. \ref{RTCT} (a). The normal state resistivity of the sample displays a typical metallic behavior. For $T$ below $\sim$ 25 K, the resistivity is well fitted by the Fermi-liquid (FL) expression, $\rho(T) = \rho_0 + AT^2$. The resultant $\rho_0$ and $A$ are 2.5 $\mu\Omega$ m and $9\times10^{-2}$ $\mu\Omega$ cm $\text{K}^{-2}$, respectively. It is worth noting, this $A$ value is almost one order of magnitude larger than iron-based superconductor BaFe$_2$(As$_{1-x}$P$_x$)$_2$ ($\sim$ $9\times10^{-3}$ $\mu\Omega$ cm $\text{K}^{-2}$ for $x \simeq 0.49$) \cite{Analytis}, three orders of magnitude larger than recently reported Dirac nodal line semimetal PtSn$_4$ ($\sim$ $2\times10^{-4}$ $\mu\Omega$ cm $\text{K}^{-2}$) \cite{Mun} and PdSn$_4$ ($\sim$ $7\times10^{-4}$ $\mu\Omega$ cm $\text{K}^{-2}$) \cite{PdSn4}, yet one order of magnitude smaller than heavy fermion superconductor CePt$_3$Si ($\sim$ 2.35 $\mu\Omega$ cm $\text{K}^{-2}$) \cite{Bauer2004}.

The specific heat plotted as $C/T$ vs $T^2$ is depicted in Fig. \ref{RTCT} (b). A jump in $C/T$ around 2.3 K also signifies a sharp transition to the superconducting state. It is noted that $C/T$ has a small residual value as $T$$\rightarrow$0, indicating the non-superconducting fraction of our sample. The non-superconducting counterpart accounts for $\sim$10\% of the sample in volume. Through fitting the $C(T)$ data above $T_c$ to the formula $C = \gamma_n T + \beta_n T^3 + \alpha_n T^5$, the Sommerfeld coefficient $\gamma_n$ representing the electron contribution is extracted as 53.1 mJ/mol $\text{K}^2$. This $\gamma_n$ value is much larger than that in BaFe$_2$(As$_{1-x}$P$_x$)$_2$ ( $\sim$ 7 mJ/mol $\text{K}^2$) and in PtSn$_4$ ($\sim$ 4 mJ/mol $\text{K}^2$), but smaller than that in CePt$_3$Si ($\sim$ 390 mJ/mol $\text{K}^2$). This large $\gamma_n$ is consistent with the enhanced $A$ coefficient, suggesting considerable density of states at the Fermi level in the normal state. The corresponding Kadowaki-Woods ratio (KWR) $A/\gamma_n^2$ is estimated to be 32 $\mu\Omega$ cm mol$^2$K$^2$J$^{-2}$. Note that this value is even larger than those found in many strongly correlated metals\cite{Jacko,Hussey}, typically $\sim$10 $\mu\Omega$ cm mol$^2$K$^2$J$^{-2}$. The transport and thermodynamic parameters for the above mentioned compounds are summarized in Table I for comparison.

The Hall effects are also studied in Fig. \ref{hall}. The Hall resistivity $\rho_{xy}$ is found to grow linearly with field up to 9 T (Fig. \ref{hall} inset). The Hall coefficient $R_H$ is calculated as the slope of $\rho_{xy}(B)$ curves. As seen, the value of $R_H$ is negative in all temperature range studied, indicating the dominant electron carriers. $R_H$ increases with the decreasing temperature before saturating at $T < 20$ K.

Fig. \ref{HP} shows the pressure ($P$) effects on La$_7$Ir$_3$. With increasing $P$, the normal state resistivity decreases slightly, and the superconducting transition temperature is progressively suppressed. As shown in Fig. \ref{HP} (b), $T_c$ is linearly suppressed with $P$. The pressure suppression rate for $T_c$ is $\sim$ -0.15 K/GPa.

Temperature sweeps and field sweeps for the superconducting transitions are shown in Figs. \ref{Hc2} (a)-(c). S1 and S2 labeled the two major samples studied in the transport measurements. The temperature is measured down to 160 mK in the dilution refrigerator. As shown in Fig. \ref{Hc2} (d), the upper critical field ($\mu_0H_{c2}$) is extracted from both $B$- and $T$- sweeps, using both 50\% and 90\% of $\rho_n$ criteria ($\rho_n$ is the normal state resistivity). Note that the abscissa of Fig. \ref{Hc2} (d) is the reduced temperature $T/T_c$. It is found that $\mu_0H_{c2}(T)$ curves from different criteria and different samples nearly collapse onto a single curve. These curves can be nicely fitted by the one-band Werthamer-Helfand-Hohenberg (WHH) model in the dirty limit. The reason of an upward curvature seen in 90\% criterion of S1 is not clear, which may be an artefact due to the impurity phase or due to an inhomogeneous current distribution arising from contact-sample interface issues. The fit gives zero-temperature $\mu_0H_{c2}$(0) close to 1.6 T, which is apparently smaller than the Pauli paramagnetic limit $H_p = 1.84T_c \simeq$ 4 T by a factor of 2-3.

\begin{figure}
\begin{center}
\includegraphics[width=12.0cm]{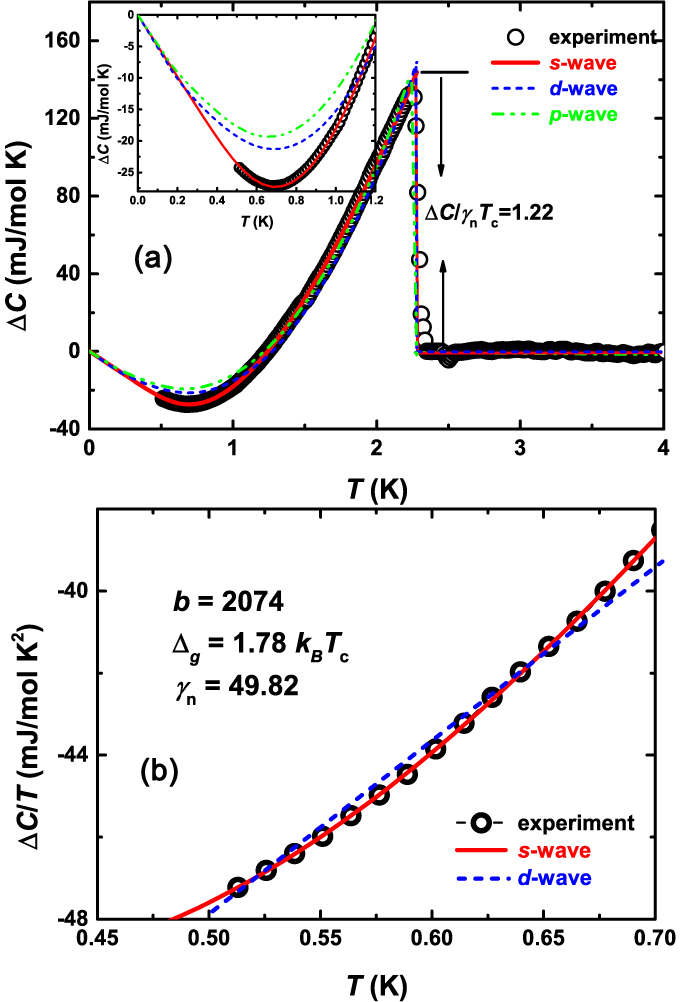}
\caption{\label{deltaC} (a) Experimental data of $\Delta C = C(0 \text{T}) - C(2 \text{T})$ vs $T$, plotted with the fits by different gap functions. In the $p$-wave fitting, we assume the gap function to be $\Delta=\Delta_0$cos$\phi$. Inset shows the expanded view of the fits at low temperatures. Clearly, $s$-wave model fits the experimental data better. The $\Delta C/\gamma_n T_c$ is 1.22, very close to the weak coupling BCS value of 1.43. (b) $\Delta C/T$ vs $T$. The solid and dashed lines are fits to the $s$-wave and $d$-wave models for $T < \frac{1}{3}T_c$, respectively.}
\end{center}
\end{figure}

We now turn to the detailed analysis of specific heat below $T_c$, which provides unambiguous evidence for the superconducting gap symmetry. In Fig. \ref{deltaC} (a), $\Delta C = C(0 \text{T}) - C(2 \text{T})$ is shown. At $\mu_0 H$= 2 T, the jump of $C$ due to superconducting transition is totally suppressed (data overlap with the red fitting line in Fig. \ref{RTCT} (b)). As the heat capacity from phonon is independent of magnetic field, $\Delta C$ excludes the phonon contribution as well as that from the nonsuperconducting part. We fit $\Delta C$ with different gap functions as was done in Ref. \cite{Carrington07,Xu13,NiuCQ}. In the BCS theory, the zero field electronic specific heat in the superconducting state is derived from the entropy $S_{es}$ which is written as

\begin{equation}
S_{es} = -\frac{3\gamma_n}{k_B \pi^3}\int_0^{2\pi}\int_0^\infty[(1-f)\ln(1-f)+f\ln f]d\varepsilon d\phi
\end{equation}

\noindent where $f$ denotes the quasiparticle occupation number $f = (1+e^{E/k_BT})^{-1}$ with $E = \sqrt{\varepsilon^2 +\Delta^2(\phi)}$. $\Delta(\phi)$ is the angle dependence of the gap function. For a conventional $s$-wave superconductor, $\Delta(\phi) = \alpha\Delta_{\text{BCS}}^s(T)$, while for a $d$-wave superconductor, $\Delta(\phi) = \alpha\Delta_{\text{BCS}}^d(T)\cos(2\phi)$. For simplicity, we assume $p$-wave gap function to be $\Delta=\Delta_0$cos$\phi$. The electronic specific heat is calculated by $C_{es} = T(\partial S/\partial T)$. As seen in Fig. \ref{deltaC} (a), at first sight, all three models fit the experimental data well. However, a blow-up view can clearly distinguish the differences among different models (see inset in Fig. \ref{deltaC} (a)). At low temperatures, both $p$-wave and $d$-wave models deviate significantly from the experimental data, while the $s$-wave model reproduces the data very well. The entropy-conserving construction at $T_c$ gives $\Delta C/\gamma_n T_c$ = 1.22, close to the weak coupling BCS value of 1.43. The $s$-wave fit gives $\gamma_n \sim$ 49 mJ/mol $\text{K}^2$, slightly smaller than the value ($\gamma_n$ $\sim$ 53.1 mJ/mol $\text{K}^2$) obtained in Fig. \ref{RTCT} (b), which means small amounts of the sample still nonsuperconducting. A rough estimate based on the ratio between these two values lead to a superconducting volume fraction around 92\%. From the $s$-wave fit, the resultant $\alpha$=1 indicates the weak coupling BCS gap.

We delve further into the temperature dependence of $\Delta C/T$ in the low-$T$ range ($T < 1/3T_c$). In this low temperature limit, one would expect $C_{es} \sim T^2$, i.e., $\Delta C/T \sim aT-\gamma_n$ for a clean $d$-wave superconductor, and $\Delta C/T \simeq bT^{-5/2}\exp(-\Delta_g/k_BT)-\gamma_n$ for the nodeless $s$-wave model. Here, $\Delta_g$ is the energy gap at zero temperature. As shown in Fig. \ref{deltaC} (b), the experimental data of $\Delta C/T$ versus $T$ clearly deviates from a linear relation, i.e., $d$-wave model is an inadequate description for La$_7$Ir$_3$. The $s$-wave model can again fit the experimental data better. The resultant $\Delta_g$ gives $\Delta_g/k_B T_c = 1.78$, which is also close to the weak coupling BCS value ($\sim 1.76$).

\begin{figure}
\begin{center}
\includegraphics[width=12.0cm]{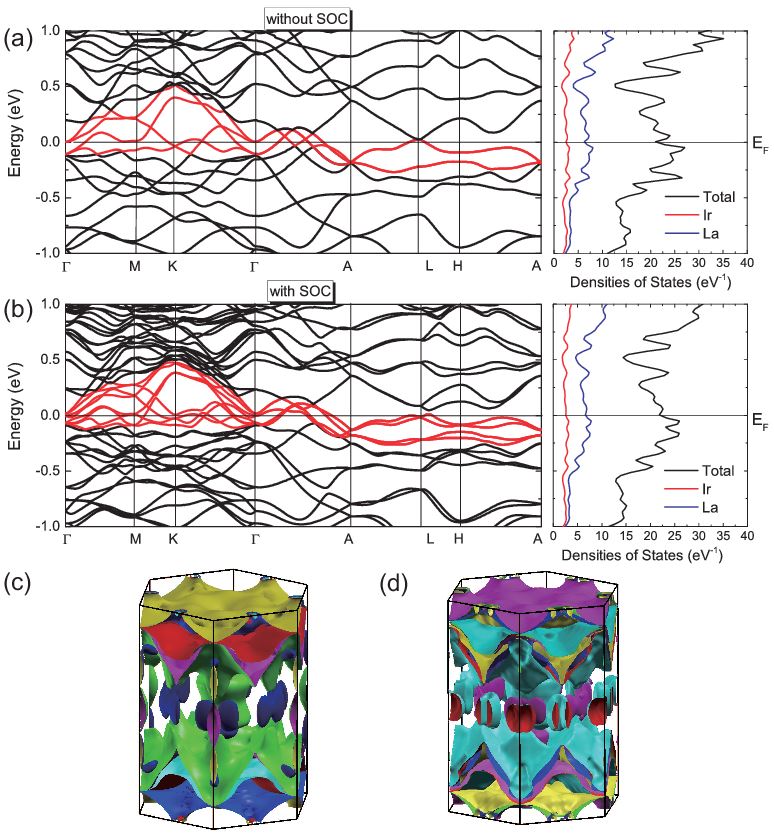}
\caption{\label{band} Calculated band structure of La$_{7}$Ir$_{3}$ (a) without SOC and (b) with SOC. Corresponding total and partial densities of states are shown in the right panels. Fermi surfaces (c) without SOC and (d) with SOC.}
\end{center}
\end{figure}

We further calculated the electronic structure and the phonon dynamics of the La$_7$Ir$_3$. As both La and Ir are relatively heavy elements, we have taken into account the spin-orbital coupling (SOC) in the calculations. For band calculations without SOC, there are four bands crossing the Fermi level ($E_{F}$) along the high symmetry path in the first Brillouin zone (BZ) as shown in Fig. \ref{band}(a). The corresponding total and partial densities of states are shown in the right panel. The bands crossing $E_{F}$ are constructed by the hybridization of La $5d$ and Ir $5d$ orbitals, and La has the dominant contribute around the Fermi level. The calculated band structure with SOC is shown in Fig. \ref{band}(b), where the number of bands doubles due to the lifting of the degeneracy. The calculated total density of states at $E_{F}$ is $\sim$ 21 eV$^{-1}$ per cell (2 formula units) with SOC. The corresponding bare specific heat coefficient $\gamma_{0}$ is 25.7 mJ/mol $\text{K}^2$. The resultant Fermi surfaces (FSs) are plotted in \ref{band}(c) and (d), without and with the inclusion of SOC respectively, showing complex three dimensional characteristics.

\begin{figure}
\begin{center}
\includegraphics[width=12.0cm]{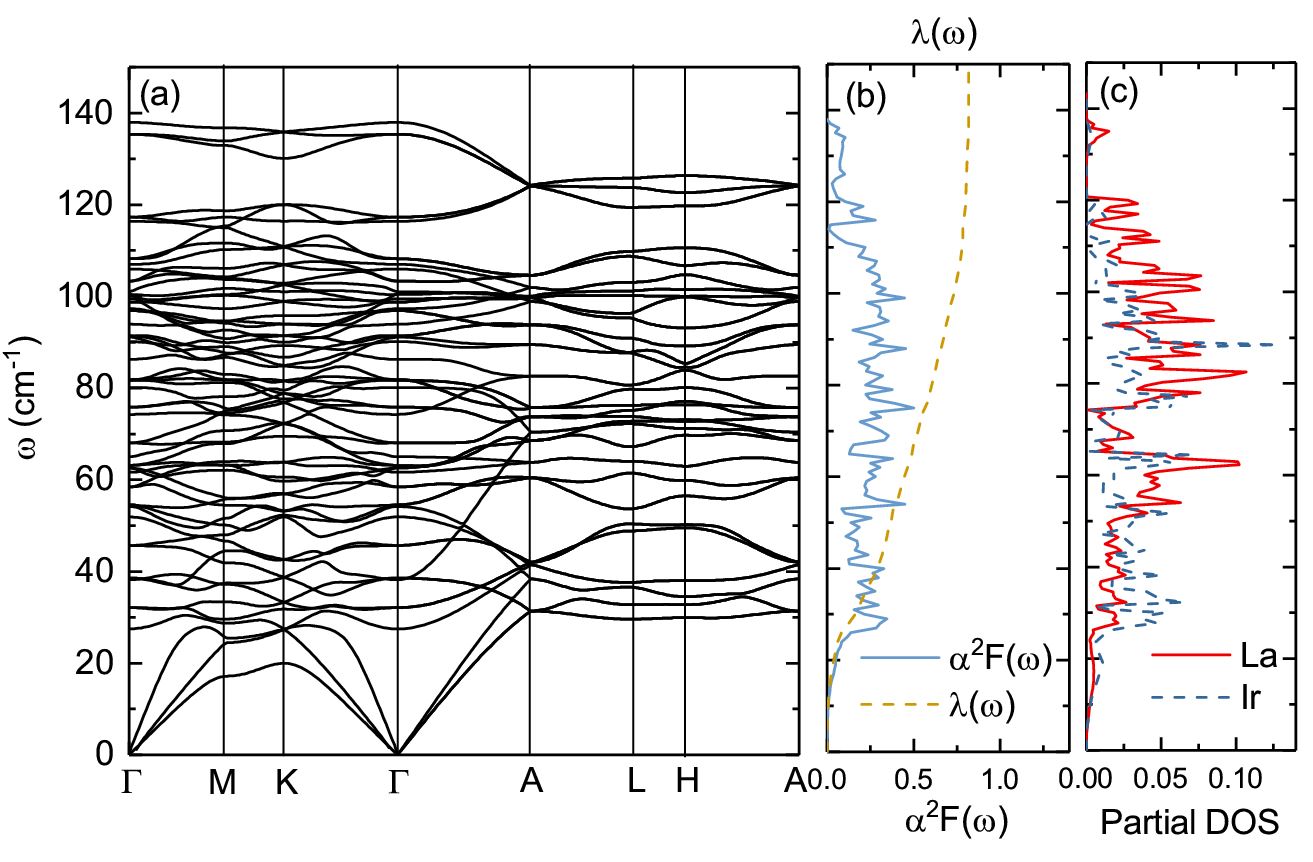}
\caption{\label{phonon} Lattice dynamics and electron-phonon properties of La$_{7}$Ir$_{3}$. Left: Phonon dispersion relations, Middle: Eliashberg function $\alpha^{2}F(\omega)$ (solid line) and frequency-dependent electron-phonon coupling $\lambda(\omega)$(dashed line). Right: Atom-projected phonon DOS.}
\end{center}
\end{figure}

Fig. \ref{phonon} summarizes the lattice dynamics properties of La$_{7}$Ir$_{3}$. In the left panel of Fig. \ref{phonon}, we show the calculated phonon dispersions. The acoustic modes are not strongly anisotropic. There are 60 phonon bands in total extending up to $\sim$ 140 cm$^{-1}$ and the point group at $\Gamma$ point is $C_{6v}$. $\Gamma$ modes can be decomposed as $\Gamma=10E_1\bigoplus10E_2\bigoplus7A_1\bigoplus3A_2\bigoplus3B_1\bigoplus7B_2$, with $E_1$ and $E_2$ modes doubly degenerate. The frequency of each mode at $\Gamma$ is listed in Table \ref{phonon-modes}.

\begin{table*}
\caption{Phonon mode frequencies (cm$^{-1}$) at $\Gamma$ point in La$_{7}$Ir$_{3}$. I: infrared active, R: Raman active.}\label{phonon-modes}
\begin{center}
\begin{tabular}{c c c c c c c c c c c c c c c c c}
\hline
&&&&&$\Gamma$(0, 0, 0)&&&&&\\
\hline
  $E_1(I+R)$& 0& 32.1& 54.6&62.6& 75.8&81.9&91.3&100.5&108.2&135.4\\
  $E_2(R)$  & 38.6&45.7&58.4&63.1&68.1&81.7&93.9&98.7&103.3&117.3 \\
  $A_1(I+R)$& 0&38.1&74.3&91.1&93.9&97.3&116.4   \\
  $A_2$     & 27.5&65.0&101.0 \\
  $B_1$     & 51.9&61.7&99.5 \\
  $B_2$     & 54.2&80.1&86.2&90.2&105.9&107.0&138.0  \\
\hline
\end{tabular}
\end{center}
\end{table*}

In the middle panel of Fig. \ref{phonon}, we present Eliashberg spectral function $\alpha^2F(\omega)$, and electron-phonon coupling $\lambda(\omega)$. In the right panel of the same figure, we show the atom-projected phonon DOS. Similarly to the electronic bands, the phonon branches have fewer dispersions in the $z$ direction. Analyzing the the phonon eigenvectors reveals that there is no clear separation between in and out-of-plane vibrations, as often happens in layered compounds. The phonon DOS distributes continuously in the frequency range up to 120 cm$^{-1}$. A small phonon gap of 10 cm$^{-1}$ occurs between 120 cm$^{-1}$ and 130 cm$^{-1}$. The projected phonon DOS shows that vibrations of La and Ir occupy the same frequency range, and the eigenvectors have a strongly mixed character. The vibration of Ir dominates the frequency range below 50 cm$^{-1}$, due to its relatively lager atom mass.

In the density-functional perturbation theoretical (DFPT)\cite{linear} calculations, the Eliashberg spectral function depends directly on the electron-phonon matrix element:
      \begin{eqnarray}
\alpha^{2}F(\omega)=\frac{1}{N(E_F)N_{k}}\sum_{kq\nu}
    \mid g^{\nu}_{n\mathbf{k},m(\mathbf{k}+\mathbf{q})}\mid ^{2}
    \times\delta(\varepsilon_{n\mathbf{k}})
    \delta(\varepsilon_{m(\mathbf{k}+\mathbf{q})})
    \delta(\omega-\omega_{q\nu}).
      \end{eqnarray}
Here, $N_{k}$ is the number of k points used in the summation, $N(E_F)$ is the density of states at the Fermi level, and $\omega_{q\nu}$ are the phonon frequencies. The electron-phonon matrix element $\mid g^{\nu}_{n\mathbf{k},m(\mathbf{k}+\mathbf{q})}\mid ^{2}$ is defined by the variation in the self-consistent crystal potential. From the $\lambda(\omega)$ curves in Fig. \ref{phonon}(b), where $\lambda(\omega)=2\int^{\omega}_{0}[\alpha^2F(\Omega)/\Omega]d\Omega$, one finds that $\lambda(100)\simeq 0.75$ is close to $\lambda(\infty)\simeq 0.81$, indicating that the phonon modes in the low and intermediate frequency regimes below 100 cm$^{-1}$ have the dominant contribution to the electron-phonon coupling. Summarizing the electron-phonon calculations, we get electron-phonon coupling $\lambda$ = 0.81 and logarithmically averaged frequency $\omega_{ln}$ = 78.5 K. Using the Allen-Dynes formula ~\cite{Allen1975,Mcmillan1968}

\begin{equation}
T_c=\frac{\omega_{ln}}{1.2}\textrm{exp}\left[-\frac{1.04(1+\lambda)}{\lambda-\mu^*(1+0.62\lambda)}\right],
\end{equation}

\noindent with the Coulomb parameter $\mu^*$= 0.15, we finally obtain $T_c$ = 2.5 K, which is close to the experimental value $T_c$ = 2.3 K. It then follows that La$_{7}$Ir$_{3}$ is very likely a conventional electron-phonon superconductor, which seems consistent with our experimental results. From the above calculated electronic specific heat coefficient $\gamma_{0}$ 25.7 mJ/mol $\text{K}^2$ and the electron-phonon coupling constant $\lambda \simeq 0.81$, we expect $\gamma_{n}$=(1+$\lambda$)$\gamma_{0}$= 46.5 mJ/mol $\text{K}^2$, close to the experimental value of 53 mJ/mol $\text{K}^2$.

\section*{Conclusion}

The transport and thermodynamic properties of La$_7$Ir$_3$ presented in our study show clear evidence for the dominant spin singlet pairing, in agreement with the $\mu$SR measurements. TRS-breaking spin triplet state usually has gap nodes (or anisotropy) in the order parameter. In broken TRS compounds LaNiC$_2$ and LaNiGa$_2$\cite{LaNiGa2-1,LaNiGa2-2}, recent study proposed a novel triplet pairing state, namely, pairing between electrons with the same spins but on different orbitals. This unusual pairing inevitably leads to two even-parity ($s$-wave) gaps\cite{Yuan16}. In our La$_7$Ir$_3$ compound, however, no evidence for the two $s$-wave gaps is observed from $H_{c2}$ and the heat capacity.

In summary, we presented the detailed physical properties of a recently reported nonnoncentrosymmetric superconductor La$_7$Ir$_3$ in which time reversal symmetry was proposed to be broken via the detection of quasi-static internal magnetic field below $T_c$. The La$_7$Ir$_3$ superconductor is found to show a notably large electronic heat capacity (Sommerfeld coefficient $\gamma_n$ $\sim$ 53 mJ/mol $\text{K}^2$) and a strongly enhanced Kadowaki-Woods ratio (KWR $\sim$ 32 $\mu\Omega$ cm mol$^2$ K$^2$ J$^{-2}$). The analysis of temperature dependent upper critical field and the electronic specific heat suggests a more likely conventional $s$-wave superconductivity in La$_7$Ir$_3$. The first-principles calculations on the electronic structure and the electron-phonon properties confirmed the experimental results.


\section*{Methods}
Polycrystalline La$_7$Ir$_3$ samples were prepared by the arc melting method as previously reported \cite{Baker2015}. The sample phase has been verified through powder x-ray diffraction measurements (Fig. \ref{XRD}). Electrical transport measurements were performed by a standard four-probe method on the PPMS-9 T system (Quantum Design). To obtain the upper critical fields, measurements on both $^3$He cooling system and a dilution refrigerator have been performed to measure the resistive superconducting transitions under field well below $T_c$. High-pressure electrical measurements were carried out on PPMS using a piston cylinder type pressure cell (type: HPC-33) with Daphne 7373 as the pressure transmission medium. The specific heat data were obtained using a relaxation method via the PPMS-9 T system equipped with a $^3$He cooling option. Thermometers and the addenda were well calibrated prior to the measurements of the sample.

The first-principles calculations of the band structure and the lattice dynamic properties were carried out based on experimental crystal structure. The electronic structure calculations with high accuracy were performed using the full-potential linearized augmented plane wave (FP$-$LAPW) method implemented in the WIEN2K code.~\cite{Wien2k}  The generalized gradient approximation (GGA)~\cite{GGA} was applied to the exchange-correlation potential calculation. The muffin tin radii were chosen to be 2.5 a.u.\ for both La and Ir. The plane-wave cutoff was defined by $RK_{max}=7.0$, where $R$ is the minimum $LAPW$ sphere radius and $K_{max}$ is the plane-wave vector cutoff. Lattice dynamic properties including phonon dispersion, phonon density of states and electron-phonon coupling were performed using the Quantum-ESPRESSO~\cite{QE} code with ultrasoft pseudopotential method and the plane wave basis. The cutoffs were chosen as 50 Ry for the wave functions and 500 Ry for the charge density. The generalized-gradient approximation of Perdew-Burke-Ernzerhof (PBE)~\cite{GGA} was used for the exchange-correlation energy function. The electronic integration was performed over a $8\times8\times8$ $k$-point mesh. Dynamical matrices and the electron-phonon interaction coefficients were calculated on a $4\times4\times4$ $q$-point grid. A dense $24\times24\times24$ grid was used for evaluating the accurate electron-phonon interaction matrices.

\section*{Acknowledgements}
The authors would like to thank Zhiqiang Mao, C. M. J. Andrew for the fruitful discussion. This work is sponsored by the National Key Basic Research Program of China (Grant No. 2014CB648400), National Natural Science Foundation of China (Grant No. 11474080, U1732162, 11504182, 11374043, 11504329, 11704047), Natural Science Foundation of Jiangsu province (Grant No. BK20150831, 15KJB140006), NUPTSF (Grant No. NY214022), Natural Science Foundation of Jiangsu Educational Department (Grant No. 15KJA430001), and six-talent peak of Jiangsu Province (Grants No. 2012-XCL-036). X. X. would also like to acknowledge the financial support from an open program from Wuhan National High Magnetic Field Center (2015KF15). B. Li, C. Q. Xu and W. Zhou contributed equally to this work.

\section*{Author contributions statement}
B.L. performed the first-principles calculations. C.Q.X. and W.Z. grew the samples and performed the most of the experiments. A.F.B. measured the resistivity in the dilution refrigerator. W.H.J., R.S., F.M.Z., H.H.H., X.F.J., B.Q., B.C. contributed to the measurements. X.F.X., B.Q. and B.L. designed and directed the project. X.F.X., B.L., and W.Z. wrote the manuscript. All authors contributed to the discussion on the results.

\section*{Additional information}
Competing financial interests: The authors declare no competing financial interests.

\begin{thebibliography}{99}
\bibitem{Yuan17} Smidman, M. \emph{et al.} Superconductivity and spin-orbit coupling in non-centrosymmetric materials: a review. \emph{Rep. Prog. Phys.} \textbf{80}, 036501 (2017).
\bibitem{Bauer2004} Bauer, E. \emph{et al.} Heavy Fermion Superconductivity and Magnetic Order in Noncentrosymmetric CePt$_{3}$Si. \emph{Phys. Rev. Lett.} \textbf{92}, 027003 (2004).
\bibitem{Bauer2005} Bauer, E., Bonalde, I. \& Sigrist, M. Superconductivity and normal state properties of non-centrosymmetric CePt$_{3}$Si: a status report. \emph{Low Temp. Phys.} \textbf{31}, 748 (2005).
\bibitem{Bauer2007} Bauer, E. \emph{et al.} Heavy Fermion Superconductivity and Antiferromagnetic Ordering in CePt$_3$Si without Inversion Symmetry, \emph{J. Phys. Soc. Jpn.} \textbf{76}, 051009 (2007).
\bibitem{Eguchi2011} Eguchi, G. \emph{et al.} Crystallographic and superconducting properties of the fully gapped noncentrosymmetric 5d-electron superconductors CaMSi$_3$, (M=Ir, Pt). \emph{Phys. Rev. B} \textbf{83}, 2385-2385 (2011).
\bibitem{Anand2011} Anand, V. K. \emph{et al.} Specific heat and $\mu$SR study on the noncentrosymmetric superconductor LaRhSi. \emph{Phys. Rev. B} \textbf{83}, 064522 (2011).
\bibitem{Smidman2014} Smidman, M. \emph{et al.} Investigations of the superconducting states of noncentrosymmetric LaPdSi$_{3}$ and LaPtSi$_{3}$. \emph{Phys. Rev. B} \textbf{89}, 094509 (2014).
\bibitem{Anand2014} Anand, V. K. \emph{et al.} Physical properties of noncentrosymmetric superconductor LaIrSi$_{3}$: A $\mu$SR study. \emph{Phys. Rev. B} \textbf{90}, 014513 (2014).
\bibitem{Togano2004} Togano, K. \emph{et al.} Superconductivity in the Metal Rich Li-Pd-B Ternary Boride. \emph{Phys. Rev. Lett.} \textbf{93}, 247004 (2004).
\bibitem{Yuan2006} Yuan, H. Q. \emph{et al.} S-Wave Spin-Triplet Order in Superconductors without Inversion Symmetry: Li$_{2}$Pd$_{3}$B and Li$_{2}$Pt$_{3}$B. \emph{Phys. Rev. Lett.} \textbf{97}, 017006 (2006).
\bibitem{LiPdPtB} Badica P. \emph{et al.} Superconductivity in a New Pseudo-Binary Li$_2$B(Pd$_{1-x}$Pt$_x$)$_3$($x$=0-1) Boride System. \emph{J. Phys. Soc. Jpn.} \textbf{74}, 1014 (2005).
\bibitem{Pecharsky1998} Pecharsky, V. K., Miller, L. L. \& Gschneidner K. A. Low-temperature behavior of two ternary lanthanide nickel carbides: Superconducting LaNiC$_2$ and magnetic CeNiC$_2$. \emph{Phys. Rev. B} \textbf{58}, 497 (1998).
\bibitem{Hillier2009} Hillier, A. D., Quintanilla, J. \& Cywinski, R. Evidence for Time-Reversal Symmetry Breaking in the Noncentrosymmetric Superconductor LaNiC$_2$. \emph{Phys. Rev. Lett.} \textbf{102}, 117007 (2009).
\bibitem{Bonalde2011} Bonalde, I. \emph{et al.} Nodal gap structure in the noncentrosymmetric superconductor LaNiC$_2$ from magnetic-penetration-depth measurements. \emph{New J. Phys.} \textbf{13}, 123022 (2011).
\bibitem{Matthias} Matthias, B. T., Compton, V. B., \& Corenzwit E. Some new superconducting compounds. \emph{J. Phys. Chem. Solids} \textbf{19}, 130 (1961).
\bibitem{Singh2014} Singh, R. P. \emph{et al.} Detection of Time-Reversal Symmetry Breaking in the Noncentrosymmetric Superconductor Re$_6$Zr Using Muon-Spin Spectroscopy. \emph{Phys. Rev. Lett.} \textbf{112}, 107002 (2014).
\bibitem{Cava14} Ali, M. N. \emph{et al.} Noncentrosymmetric superconductor with a bulk three-dimensional Dirac cone gapped by strong spin-orbit coupling. \emph{Phys. Rev. B} \textbf{89}, 020505(R) (2014).
\bibitem{Xu_PbTaSe} Xu, C. Q. \emph{et al.} Topological phase transition under pressure in the topological nodal-line superconductor PbTaSe$_2$. \emph{Phys. Rev. B} \textbf{96}, 064528 (2017).
\bibitem{Baker2015} Barker, J. A. \emph{et al.} Unconventional Superconductivity in La$_7$Ir$3$ Revealed by Muon Spin Relaxation: Introducing a New Family of Noncentrosymmetric Superconductor That Breaks Time-Reversal Symmetry. \emph{Phys. Rev. Lett.} \textbf{115}, 267001 (2015).
\bibitem{Biswas} Biswas P. K. \emph{et al.} Evidence for superconductivity with broken time-reversal symmetry in locally noncentrosymmetric
SrPtAs. \emph{Phys. Rev. B} \textbf{87}, 180503 (2013).
\bibitem{Analytis} Analytis, J. G. \emph{et al.} Transport near a quantum critical point in BaFe$_2$(As$_{1-x}$P$_x$)$_2$. Nat. Phys. \textbf{10}, 194-197 (2014)
\bibitem{Mun} Mun, E. \emph{et al.} Magnetic field effects on transport properties of PtSn$_4$. \emph{Phys. Rev. B} \textbf{85}, 035135 (2012).
\bibitem{PdSn4} Xu, C. Q. \emph{et al.} Enhanced electron correlations in the binary stannide PdSn$_4$: A homologue of the Dirac nodal arc semimetal PtSn$_4$. \emph{Phys. Rev. Mater.} \textbf{1}, 064201 (2017).
\bibitem{Jacko} Jacko A. C. \emph{et al.} A unified explanation of the Kadowaki-Woods ratio in strongly correlated metals. Nat. Phys. \textbf{5}, 422 (2009).
\bibitem{Hussey} Hussey, N. E. Non-generality of the Kadowaki-Woods ratio in correlated oxides. J. Phys. Soc. Jpn. \textbf{74}, 1107-1110 (2005).
\bibitem{Carrington07} Taylor, O. J., Carrington, A., Schlueter, J. A. Specific-Heat Measurements of the Gap Structure of the Organic Superconductors $\kappa$-(ET)$_2$Cu[N(CN)$_2$]Br and $\kappa$-(ET)$_2$Cu(NCS)$_2$. \emph{Phys. Rev. Lett.} \textbf{99}, 057001 (2007).
\bibitem{Xu13} Xu, Xiaofeng \emph{et al.} Evidence for two energy gaps and Fermi liquid behavior in the SrPt$_2$As$_2$ superconductor. \emph{Phys. Rev. B} \textbf{87}, 224507 (2013).
\bibitem{NiuCQ} Niu, C. Q. \emph{et al.} Effect of selenium doping on the superconductivity of Nb$_2$Pd(S$_{1-x}$Se$_x$)$_5$. \emph{Phys. Rev. B} \textbf{88}, 104507 (2013).
\bibitem{linear} Baroni, S. \emph{et al.} Phonons and related crystal properties from density-functional perturbation theory. \emph{Rev. Mod. Phys.} \textbf{73}, 515 (2001).
\bibitem{Allen1975} Allen, P. \& Dynes, R. Transition temperature of strong-coupled superconductors reanalyzed. \emph{Phys. Rev. B} \textbf{12}, 905 (1975).
\bibitem{Mcmillan1968} Mcmillan, W. Transition Temperature of Strong-Coupled Superconductors. \emph{Phys. Rev.} \textbf{167}, 331 (1968).
\bibitem{LaNiGa2-1} Zeng, N. L. \& Lee, W. H. Superconductivity in the Ni-based ternary compound LaNiGa$_2$. \emph{Phys. Rev. B} \textbf{66}, 092503 (2002).
\bibitem{LaNiGa2-2} Hillier A. D. \emph{et al.} Nonunitary Triplet Pairing in the Centrosymmetric Superconductor LaNiGa$_2$. \emph{Phys. Rev. Lett.} \textbf{109}, 097001 (2012).
\bibitem{Yuan16} Weng, Z. F. \emph{et al.} Two-Gap Superconductivity in LaNiGa$_2$ with Nonunitary Triplet Pairing and Even Parity Gap Symmetry. \emph{Phys. Rev. Lett.} \textbf{117}, 027001 (2016).
\bibitem{Wien2k} Blaha, P. \emph{et al.} WIEN2k, An Augmented Plane Wave + LO Program for Calculating Crystal Properties, TU Wien, Vienna, (2001).
\bibitem{GGA} Perdew, J. P., Burke, K. \& Ernzerhof, M. Generalized Gradient Approximation Made Simple. \emph{Phys. Rev. Lett.} \textbf{77}, 3865 (1996).
\bibitem{QE} Giannozzi P. \emph{et al}. QUANTUM ESPRESSO: a modular and open-source software project for quantum simulations of materials. \emph{J. Phys.: Condens. Matter} \textbf{21}, 395502 (2009).








\end{thebibliography}
\end{document}